\documentclass[preprint,showpacs,preprintnumbers,amsmath,amssymb]{revtex4}
\usepackage{epsfig}
\usepackage{graphicx}
\usepackage{dcolumn}
\usepackage{bm}
\usepackage{threeparttable}

\def\beq{\begin{equation}}
\def\eeq{\end{equation}}
\def\eeqn{\end{equation}}
\newcommand\iden{\leavevmode\hbox{\small1\normalsize\kern-.33em1}}

\newcommand{\bea} {\begin{eqnarray}}
\newcommand{\eea} {\end{eqnarray}}

\let\jnfont=\rm
\def\NPB#1 {{\jnfont Nucl.\ Phys.\ B }{\bf #1} }
\def\PLB#1 {{\jnfont Phys.\ Lett.\ B }{\bf #1} }
\def\EPJC#1 {{\jnfont Eur.\ Phys.\ Jour.\ C }{\bf #1} }
\def\PRD#1 {{\jnfont Phys.\ Rev.\ D }{\bf #1} }
\def\PRL#1 {{\jnfont Phys.\ Rev.\ Lett.\ }{\bf #1} }
\def\MPLA#1 {{\jnfont Mod.\ Phys.\ Lett.\ A }{\bf #1} }
\def\JPG#1 {{\jnfont J.\ Phys.\ G }{\bf #1} }
\def\CTP#1 {{\jnfont Commun.\ Theor.\ Phys.\ }{\bf #1} }
\def\JHEP#1 {{\jnfont JHEP \ }{\bf #1} }
\def\NPPS#1 {{\jnfont Nucl.\ Phys.\ Proc.\ Suppl.\ }{\bf #1} }
\def\CPC#1 {{\jnfont Comput.\ Phys.\ Commun.\ }{\bf #1} }
\def\CPL#1 {{\jnfont Chin.\ Phys.\ Lett. }{\bf #1} }
\def\APPB#1 {{\jnfont Acta\ Phys.\ Polon.\ B }{\bf #1} }

\def\lsim{\raise0.3ex\hbox{$<$\kern-0.75em\raise-1.1ex\hbox{$\sim$}}}
\def\gsim{\raise0.3ex\hbox{$>$\kern-0.75em\raise-1.1ex\hbox{$\sim$}}}
\def\PR#1 {{\jnfont Phys.\ Rept. }{\bf #1} }
\def\CHC#1 {{\jnfont Chin.\ Phys.\ C }{\bf #1} }
\def\NIMA#1 {{\jnfont Nucl.\ Instrum.\ Meth.\ A }{\bf #1} }
\def\JCAP#1 {{\jnfont JCAP \ }{\bf #1} }
\def\ASA#1 {{\jnfont Astron.\ Astrophys.\ A }{\bf #1} }  

\begin{document}

\title{\ \\[10mm] Wrong sign Yukawa
coupling of the 2HDM plus a singlet scalar dark matter confronted with
dark matter and Higgs data}

\author{Lei Wang, Rongle Shi, Xiao-Fang Han}

\affiliation{ Department of Physics, Yantai University, Yantai
264005, China}


\begin{abstract}
In the framework of type-II two-Higgs-doublet model with a scalar dark matter ($S$),
we examine the wrong sign Yukawa coupling of the 125 GeV Higgs which is the only portal between 
the dark matter and SM sectors. After imposing the constraints from the Higgs searches at the LHC
and dark matter experiments, we obtain some interesting observables:
(i) The theory, oblique parameters, and the Higgs searches at the LHC
can impose stringent constraints on $\tan\beta$ in the case of the wrong sign Yukawa coupling of the 125 GeV Higgs. 
For example, for $m_H=600$ GeV and 140 GeV $<m_A<200$ GeV, 
$\tan\beta$ is required to be in the range of 4.2 and 5.7. 
(ii) Due to the contribution of $SS\to AA$ annihilation channel to the relic density, the dark matter coupling with
the 125 GeV Higgs can be sizably suppressed. However, the $SS\to AA$ channel can not
solve the tension between the relic density and the signal data of 125 GeV Higgs for $m_S < 50$ GeV.  
(iii) The limits of XENON1T (2017) and PandaX-II (2017) exclude 
most of samples in the ranges of 65 GeV $<m_S<$ 78 GeV, $0.8<f^n/f^p<1$,
and $y_d/y_u>-0.82$. 
(iv) The Fermi-LAT limits exclude most of samples in the range of 62.5 GeV $<m_S <65$ GeV, including
the samples with $f^n/f^p\sim -0.7$.
\end{abstract}

\maketitle

\section{Introduction}
The two-Higgs-doublet model (2HDM) \cite{2hdm} is a simple extension of SM by
adding a second $SU(2)_L$ Higgs doublet, which includes two neutral CP-even Higgs bosons $h$ and
$H$, one neutral pseudoscalar $A$, and two charged Higgs $H^{\pm}$ in the scalar sector.
According to different Yukawa couplings, there are four types
of 2HDMs without the tree-level flavor changing neutral
currents, type-I \cite{i-1,i-2}, type-II \cite{i-1,ii-2},
lepton-specific, and flipped models \cite{xy-1,xy-2,xy-3,xy-4}.

In the 2HDM, the 125 GeV Higgs is allowed to have the SM-like coupling and
the wrong sign Yukawa coupling. For the former, the tree-level couplings of 
the 125 GeV Higgs and the SM particles are very close to the SM couplings. 
For the latter, compared to the SM, at least one of the Yukawa couplings of the 125 GeV Higgs
has an opposite sign to the coupling of gauge boson \cite{ws-1,ws-2,ws-3,ws-5,ws-6,ws-7,ws-8,ws-9,ws-10,ws-11,wsbu}.
The wrong sign Yukawa coupling of the 125 GeV Higgs is a characteristic of 2HDM, which has some 
interesting applications. For example, in the lepton-specific model the muon g-2 anomaly can be explained by
 a light pseudoscalar with a very large $\tan\beta$, and the corresponding 125 GeV Higgs is favored to have the 
wrong sign Yukawa coupling of lepton \cite{mug2,mug22}. Besides, in the type-II 2HDM with a scalar 
dark matter (DM), the isospin-violating DM interactions with nucleons can
be obtained when the mediator is the 125 GeV Higgs with the wrong sign Yukawa coupling of down-type quark 
\cite{2hisos-1,2hisos-2,2hisos-3,2hisos-4,2hisos-5,2hisos-6,dmbu}.
With the rapidly improved sensitivity of DM direct detection experiments,
the LUX (2016) \cite{lux}, XENON1T (2017) \cite{xenon} and PandaX-II (2017) \cite{pandax} 
impose stringent constraints on the cross section of DM-nucleon. 
If the 125 GeV Higgs does not have the wrong sign Yuakwa coupling, 
the tree-level couplings of the 125 GeV Higgs and the SM particles are required to be 
very close to the SM couplings by the experimental data. Such a 125 GeV Higgs mediates the DM interactions
with SM particles, the cross section for DM scattering off a nucleus in the 2HDM with a scalar DM is approximately
the same as that in the SM with a scalar DM. If DM only annihilates into the SM particles,
the LUX (2016) and PandaX-II (2016) limits exclude the DM mass up to 330 GeV, except a small range near
the resonance point $m_{DM}=m_h/2$ in the 2HDM with a scalar DM \cite{2hisos-6}. 
However, in the case of the wrong sign Yukawa coupling of the 125 GeV Higgs, 
the cross section for DM scattering off a nucleus will be sizably suppressed so that many
parameter space can be saved.
In addition, the annihilation channels of DM are strongly limited by the Fermi-LAT search for DM annihilation from 
dwarf spheroidal satellite galaxies (dSphs) \cite{fermi}. 
A lot of ATLAS and CMS searches for scalars at the LHC can impose the 
strong constraints on the parameter space of type-II 2HDM in the case of the wrong sign Yukawa coupling of the 125 GeV Higgs.
In this paper we will examine the wrong sign Yukawa coupling of the 125 GeV Higgs
in the type-II 2HDM with a scalar DM considering the
joint constraints from the theory, precision electroweak data,
flavor observables, the 125 GeV Higgs signal data, and the searches
for the additional Higgs at the LHC as well as DM experiments.
Compared to recent study in Ref. \cite{2hisos-6}, our work includes some new studies in addition to using the 
latest experimental data of Higgs and dark matter: (i) We include the constraints from the searches for the additional
Higgses at the LHC, which can indirectly reduce the parameter space of the wrong sign Yukawa coupling by constraining
$\tan\beta$ and $\sin(\beta-\alpha)$. (ii) The DM annihilation into a pair of pseudoscalars is considered, and the annihilation
channel can weaken the constraints of the DM experiments sizably.

Our work is organized as follows. In Sec. II we recapitulate the type-II
2HDM with a scalar DM. In Sec. III we perform
numerical calculations. In Sec. IV, we discuss the allowed parameter space
after imposing the relevant theoretical and
experimental constraints. Finally, we give our
conclusion in Sec. V.

\section{Type-II two-Higgs-doublet model with a scalar dark matter}
\subsection{Type-II two-Higgs-doublet model}
In the type-II 2HDM with a scalar DM, 
the scalar potential consists of two parts, $\mathcal{V}_{2HDM}$ and
$\mathcal{V}_{S}$. They are respectively
the original potential of type-II 2HDM and the potential of DM sector, and
$\mathcal{V}_{2HDM}$ is given by \cite{2h-poten}
\begin{eqnarray} \label{V2HDM} \mathcal{V}_{2HDM} &=& m_{11}^2
(\Phi_1^{\dagger} \Phi_1) + m_{22}^2 (\Phi_2^{\dagger}
\Phi_2) - \left[m_{12}^2 (\Phi_1^{\dagger} \Phi_2 + \rm h.c.)\right]\nonumber \\
&&+ \frac{\lambda_1}{2}  (\Phi_1^{\dagger} \Phi_1)^2 +
\frac{\lambda_2}{2} (\Phi_2^{\dagger} \Phi_2)^2 + \lambda_3
(\Phi_1^{\dagger} \Phi_1)(\Phi_2^{\dagger} \Phi_2) + \lambda_4
(\Phi_1^{\dagger}
\Phi_2)(\Phi_2^{\dagger} \Phi_1) \nonumber \\
&&+ \left[\frac{\lambda_5}{2} (\Phi_1^{\dagger} \Phi_2)^2 + \rm
h.c.\right].
\end{eqnarray}
We consider the case of the CP-conserving in which all $\lambda_i$ and
$m_{12}^2$ are real.
The two complex Higgs doublets have hypercharge $Y = 1$,
\begin{equation}
\Phi_1=\left(\begin{array}{c} \phi_1^+ \\
\frac{1}{\sqrt{2}}\,(v_1+\phi_1^0+ia_1)
\end{array}\right)\,, \ \ \
\Phi_2=\left(\begin{array}{c} \phi_2^+ \\
\frac{1}{\sqrt{2}}\,(v_2+\phi_2^0+ia_2)
\end{array}\right).
\end{equation}
Where $v_1$ and $v_2$ are the electroweak vacuum expectation values
(VEVs) with $v^2 = v^2_1 + v^2_2 = (246~\rm GeV)^2$. The ratio of the two VEVs is usually defined
as $\tan\beta=v_2 /v_1$. There are five physical states after spontaneous electroweak
symmetry breaking: two neutral
CP-even $h$ and $H$, one neutral pseudoscalar $A$, and two charged
scalars $H^{\pm}$.

The Yukawa interactions of type-II 2HDM are given by
 \bea
- {\cal L} &=&Y_{u2}\,\overline{Q}_L \, \tilde{{ \Phi}}_2 \,u_R
+\,Y_{d1}\,
\overline{Q}_L\,{\Phi}_1 \, d_R\, + \, Y_{\ell 1}\,\overline{L}_L \, {\Phi}_1\,e_R+\, \mbox{h.c.}\,, \eea where
$Q_L^T=(u_L\,,d_L)$, $L_L^T=(\nu_L\,,l_L)$,
$\widetilde\Phi_{1,2}=i\tau_2 \Phi_{1,2}^*$, and $Y_{u2}$,
$Y_{d1}$ and $Y_{\ell 1}$ are $3 \times 3$ matrices in family
space.

The Yukawa couplings of the neutral Higgs bosons with respect to the SM are given by
\bea\label{hffcoupling} &&
y^{h}_V=\sin(\beta-\alpha),~~~y_{f}^{h}=\left[\sin(\beta-\alpha)+\cos(\beta-\alpha)\kappa_f\right], \nonumber\\
&&y^{H}_V=\cos(\beta-\alpha),~~~y_{f}^{H}=\left[\cos(\beta-\alpha)-\sin(\beta-\alpha)\kappa_f\right], \nonumber\\
&&y^{A}_V=0,~~~y_{A}^{f}=-i\kappa_f~{\rm (for~u)},~~~~y_{f}^{A}=i \kappa_f~{\rm (for~d,~\ell)},\nonumber\\ 
&&{\rm with}~\kappa_d=\kappa_\ell\equiv-\tan\beta,~~~\kappa_u\equiv 1/\tan\beta,\eea 
where $\alpha$ is the mixing angle of the two CP-even Higgs bosons, and $V$ denotes $Z$ or $W$.

In the SM, the Yukawa couplings of the 125 GeV Higgs to the fermion pairs have the same sign as the corresponding
coupling of gauge boson. In the model, the Yukawa coupling of the 125 GeV Higgs
can be opposite in sign to the coupling to gauge boson, and such Yukawa coupling is defined 
as the wrong-sign Yukawa coupling. However, the 125 GeV Higgs signal data
require the absolute values of the Higgs couplings to be close to the SM.
Therefore, we can obtain
\begin{align}  &y_h^{f_i}=-1+\epsilon,~~y^{V}_h\simeq 1-0.5\cos^2(\beta-\alpha) ~~{\rm for}~ \sin(\beta-\alpha) >0~{\rm and}~\cos(\beta-\alpha) >0~,\nonumber\\
& y_h^{f_i}=1-\epsilon,~~y^{V}_h\simeq -1+0.5\cos^2(\beta-\alpha) ~~{\rm for}~ \sin(\beta-\alpha)<0~{\rm and}~\cos(\beta-\alpha) >0. \end{align}
Where $\mid\epsilon\mid$ and $\mid\cos(\beta-\alpha)\mid$ are much smaller than 1.
From Eq. (\ref{hffcoupling}), we can obtain
\begin{align}\label{wrcp}
&\kappa_f=\frac{-2+\varepsilon+0.5\cos(\beta-\alpha)^2}{\cos(\beta-\alpha)}<<-1 ~{\rm for}~ \sin(\beta-\alpha) >0~{\rm and}~\cos(\beta-\alpha) >0~,\nonumber\\
&\kappa_f=\frac{2-\varepsilon-0.5\cos(\beta-\alpha)^2}{\cos(\beta-\alpha)} >>1 ~{\rm for}~ \sin(\beta-\alpha) <0~{\rm and}~\cos(\beta-\alpha) >0~.
\end{align}
In the type-II 2HDM, the charged Higgs coupling to the top quark is proportional 
to $\kappa_u$ ($\equiv1/\tan\beta$), which can give contributions to $B\to X_s\gamma$, and $\Delta m_{B_s}$, $\Delta m_{B_d}$ and
$R_b$. The limits of $B$-meson observables require the lower bound of $\tan\beta$ to increase with
decreasing of $m_{H^{\pm}}$, such as $\tan\beta>$ 0.8 ($0<\kappa_u<1.25$) for $m_{H^{\pm}}<850$ GeV \cite{1706.07414}.
Therefore, according to Eq. (\ref{wrcp}), it is very difficult to obtain the wrong sign Yukawa coupling
for the up-type quark in the model. In addition, for the $h\to \gamma\gamma$, 
the top quark loop has destructive interferences with the $W$ boson loop in the SM. Conversely,
the top quark loop will interference constructively with the $W$ boson loop in the case of
the wrong-sign Yukawa coupling of top quark in the model, and lead the width of $h\to \gamma\gamma$  
to deviate from the SM prediction sizably, which is disfavored by the diphoton signal data of the 125 GeV Higgs
\cite{1306.2941}. Therefore, in this model the wrong sign Yukawa coupling for the up-type quark is generally disfavored.
Due to $\kappa_d=\kappa_\ell\equiv-\tan\beta$ in the type-II 2HDM,  $\kappa_d>>$ 1 ($\kappa_\ell>>$ 1) is absolutely impossible, 
and $\kappa_d <<$ -1 ($\kappa_\ell << $ -1) is allowed by the limits of $B$-meson observables.
 Therefore, according to Eq. (\ref{wrcp}), there may be the wrong sign Yukawa couplings of the down-type quark and lepton only
for $\sin(\beta-\alpha) > 0$ and $\cos(\beta-\alpha) >0$. Note that in the discussion above, we take a convention \cite{2hc-1}, 
0$\leq\beta\leq \frac{\pi}{2}$ and $-\frac{\pi}{2}\leq\beta-\alpha\leq \frac{\pi}{2}$,
which leads to $0 \leq \cos(\beta-\alpha) \leq 1$ and  $-1 \leq \sin(\beta-\alpha) \leq 1$.
The physical results in the convention are equivalent to those in another convention,
$-1 \leq \cos(\beta-\alpha) \leq 1$ and $0 \leq \sin(\beta-\alpha) \leq 1$.

\subsection{A scalar dark matter}
 Now we add a real singlet scalar $S$ to the type-II 2HDM, and $S$ is a possible DM candidate. 
The potential containing the DM is given by
\begin{eqnarray}
\mathcal{V}_{S}&=&{1\over 2}S^2(\kappa_{1}\Phi_1^\dagger \Phi_1
+\kappa_{2}\Phi_2^\dagger \Phi_2)+{m_{0}^2\over
2}S^2+{\lambda_S\over 4!}S^4\label{potent}.
\end{eqnarray} The linear and cubic terms of the $S$ field are
forbidden by a $Z'_2$ symmetry $S\rightarrow -S$. The DM mass and
the cubic interactions with the neutral Higgses are obtained from the Eq.
(\ref{potent}),
\begin{eqnarray}
m_S^2&=&m_0^2+\frac{1}{2}\kappa_1
v^2\cos^2\beta+\frac{1}{2}\kappa_2 v^2\sin^2\beta,\nonumber\\
-\lambda_{h} vS^2h/2&\equiv& -(-\kappa_{1}\sin\alpha\cos\beta+\kappa_{2}\cos\alpha\sin\beta)vS^2h/2,\nonumber\\
-\lambda_{H} vS^2H/2&\equiv&
-(\kappa_{1}\cos\alpha\cos\beta+\kappa_{2}\sin\alpha\sin\beta)vS^2H/2.
\label{dmcoup}\end{eqnarray}

\section{Numerical calculations}
In our discussions we take the light CP-even Higgs boson $h$ as the
SM-like Higgs, $m_h=125$ GeV. In the model both $h$ and $H$ can be portals between
the DM and SM sectors. In Ref. \cite{2hisos-4}, the authors studied the general situation in which $h$ and $H$ may contribute to DM
interactions with SM particles.  Ref. \cite{2hisos-6} claimed that if the heavier CP-even Higgs boson as the only portal,
much of the region of DM mass below 100 GeV are excluded, and the theoretical constraints from perturbativity, vacuum stability, and unitarity play an important role \cite{2hisos-6}. In this paper we choose $\lambda_{H}=0$, and focus on 
examining the 125 GeV Higgs with the wrong sign Yukawa coupling of down-type quark as the only portal in
light of the following considerations: (i) The current signal data of the 125 GeV Higgs require the $hb\bar{b}$ coupling
to have a magnitude close to the SM value, but as a simple and interesting new physics scenario,
the wrong sign Yukawa coupling of bottom quark is still allowed. (ii) The searches for additional Higgses at the LHC can put constraints
on $\tan\beta$ and $\sin(\beta-\alpha)$, and further reduce the parameter space of wrong sign Yukawa coupling
of the 125 GeV Higgs. (iii) When the mediator is the 125 GeV Higgs with the wrong sign Yukawa coupling of down-type quark,
the isospin-violating DM interactions with nucleons can be obtained, leading to a very suppressed
cross section for DM scattering off a nucleus. Here we will examine the 
125 GeV Higgs with the wrong sign Yukawa coupling of down-type quark using the latest data of the 125 GeV Higgs signal, the
searches for additional Higgses at the LHC, and the DM experiments. Note that $\lambda_H=0$ is a choice 
imposed on the parameter space, which does not cause a new symmetry of the potential.

The measurement of the branching ratio of $b \to s\gamma$ imposed a
strong lower limit on the charged Higgs mass of type-II 2HDM, $m_{H^{\pm}} > 570$ GeV \cite{bsr570}.
The LHC searches for the charged scalar fail to constrain the
model for $m_{H^{\pm}}>500$ GeV \cite{mhp500}.
Our previous paper shows that the $S$, $T$ and $U$ oblique parameters give the strong
constraints on the mass spectrum of Higgses \cite{1701.02678}. 
One of $m_A$ and $m_H$ is around 600 GeV, another is allowed to have 
a wide mass range including low mass. Thus, we take $m_H=600$ GeV and $m_A>$ 20 GeV. For such case, the $SS\to AA$ annihilation 
channel is kinematically open, and gives an important contribution to the DM relic density.
If the $SS\to AA$ annihilation channel is fully responsible for the current relic density, 
the average cross section of the annihilation at present time can suffer from constraints of
the Fermi-LAT search for DM annihilation from dSphs. 
In order to avoid the tension, we take $S$ to be lighter than $A$.

In our calculation, we consider the following observables and constraints:

\begin{itemize}
\item[(1)] Theoretical constraints. In the model, the scalar potentials include the 
original potential type-II 2HDM and the potential of DM sector. The parameters suffer from
the constraints of the vacuum stability, perturbativity, and tree-level unitarity, which are
discussed in detail in Refs. \cite{2hisos-4,2hisos-6}. Here we follow the formulas in \cite{2hisos-4,2hisos-6} to
perform the theoretical constraints. Note that there are additional factors of $\frac{1}{2}$ in $\kappa_1$ 
term and $\kappa_2$ term of this paper compared to Refs. \cite{2hisos-4,2hisos-6}.

\item[(2)] Oblique parameters. The $S$, $T$, $U$ parameters can give stringent constraints on 
the mass spectrum of Higgses of 2HDM. The $\textsf{2HDMC}$ \cite{2hc-1}
is used to consider the constraints from
the oblique parameters ($S$, $T$, $U$).

\item[(3)] The flavor observables and $R_b$. We include the constraints of $B$-meson
decays from $B\to X_s\gamma$, $\Delta m_{B_s}$ and $\Delta m_{B_d}$.  $\textsf{SuperIso-3.4}$ \cite{spriso} is
employed to calculate $B\to X_s\gamma$, and $\Delta m_{B_s}$ and $\Delta m_{B_d}$ are calculated following the
formulas in \cite{deltmq}. Besides, we implement the constraints of bottom quarks produced in $Z$ decays, $R_b$,
which is calculated using the formulas in \cite{rb1,rb2}.

\item[(4)] The global fit to the signal data of the 125 GeV Higgs. 
In this model, the 125 GeV Higgs couplings with the SM particles
can be modified compared to the SM, which can give the corrections to
the SM-like decay modes. In addition, if kinematically allowed, $h\to AA$ and $h\to SS$
modes can open, and enhance the total width of $h$ sizably, which will
be strongly constrained by the 125 GeV Higgs data. We perform the
$\chi^2$ calculation for the signal strengths of the 125 GeV Higgs in the
$\mu_{ggF+tth}(Y)$ and $\mu_{VBF+Vh}(Y)$ with $Y$ denoting the decay
modes $\gamma\gamma$, $ZZ$, $WW$, $\tau^+ \tau^-$ and $b\bar{b}$,
 \begin{eqnarray} \label{eq:ellipse}
  \chi^2(Y) =\left( \begin{array}{c}
        \mu_{ggH+ttH}(Y) - \widehat{\mu}_{ggH+ttH}(Y)\\
        \mu_{VBF+VH}(Y) - \widehat{\mu}_{VBF+VH}(Y)
                 \end{array} \right)^T
                 \left(\begin{array}{c c}
                        a_Y & b_Y \\
                        b_Y & c_Y
                 \end{array}\right) \nonumber\\
\times
                  \left( \begin{array}{c}
        \mu_{ggH+ttH}(Y) - \widehat{\mu}_{ggH+ttH}(Y)\\
        \mu_{VBF+VH}(Y) - \widehat{\mu}_{VBF+VH}(Y)
                 \end{array} \right) \,.
 \end{eqnarray}
$\widehat{\mu}_{ggH+ttH}(Y)$ and $\widehat{\mu}_{VBF+VH}(Y)$
are the best-fit values and $a_Y$, $b_Y$ and $c_Y$ are the
parameters of the ellipse. These parameters are given by the
combined ATLAS and CMS experiments \cite{160602266}. We pay
particular attention to the surviving samples with
$\chi^2-\chi^2_{\rm min} \leq 6.18$, where $\chi^2_{\rm min}$
denotes the minimum of $\chi^2$. These samples correspond to be within
the $2\sigma$ range in any two-dimension plane of the
model parameters when explaining the Higgs data.

\begin{table}
\begin{footnotesize}
\begin{tabular}{| c | c | c | c |}
\hline
\textbf{Channel} & \textbf{Experiment} & \textbf{Mass range (GeV)}  &  \textbf{Luminosity} \\
\hline
 {$gg/b\bar{b}\to A/H \to \tau^{+}\tau^{-}$} & ATLAS 8 TeV~\cite{47Aad:2014vgg} & 90-1000 & 19.5-20.3 fb$^{-1}$ \\
{$gg/b\bar{b}\to A/H \to \tau^{+}\tau^{-}$} & CMS 8 TeV~\cite{48CMS:2015mca} &  90-1000  &19.7 fb$^{-1}$ \\
 {$b\bar{b}\to A/H \to \tau^{+}\tau^{-}$} & CMS 8 TeV \cite{1511.03610}& 25-80   & 19.7 fb$^{-1}$ \\
{$gg/b\bar{b}\to A/H \to \tau^{+}\tau^{-}$} & ATLAS 13 TeV~\cite{82vickey} & 200-1200 &13.3 fb$^{-1}$ \\
{$gg/b\bar{b}\to A/H \to \tau^{+}\tau^{-}$} & CMS 13 TeV~\cite{add-hig-16-037} & 90-3200 &12.9 fb$^{-1}$ \\
\hline
 {$gg\to h \to AA \to \tau^{+}\tau^{-}\tau^{+}\tau^{-}$} & ATLAS 8 TeV~\cite{1505.01609} & 4-50 & 20.3 fb$^{-1}$ \\
{$pp\to  h \to AA \to \tau^{+}\tau^{-}\tau^{+}\tau^{-}$} & CMS 8 TeV~\cite{1701.02032} &  5-15  &19.7 fb$^{-1}$ \\
{$pp\to  h \to AA \to (\mu^{+}\mu^{-})(b\bar{b})$} & CMS 8 TeV~\cite{1701.02032} &  25-62.5  &19.7 fb$^{-1}$ \\
{$pp\to  h \to AA \to (\mu^{+}\mu^{-})(\tau^{+}\tau^{-})$} & CMS 8 TeV~\cite{1701.02032} &  15-62.5  &19.7 fb$^{-1}$ \\
\hline
$gg\to A\to hZ \to (\tau^{+}\tau^{-}) (\ell \ell)$ & CMS 8 TeV \cite{66Khachatryan:2015tha}& 220-350 & 19.7 fb$^{-1}$\\

$gg\to A\to hZ \to (b\bar{b}) (\ell \ell)$ & CMS 8 TeV \cite{67Khachatryan:2015lba} & 225-600 &19.7 fb$^{-1}$ \\

$gg\to A\to hZ\to (\tau^{+}\tau^{-}) Z$ & ATLAS 8 TeV \cite{68Aad:2015wra}&220-1000 & 20.3 fb$^{-1}$ \\

 {$gg\to A\to hZ\to (b\bar{b})Z$} & ATLAS 8 TeV \cite{68Aad:2015wra}& 220-1000 & 20.3 fb$^{-1}$  \\
{$gg/b\bar{b}\to A\to hZ\to (b\bar{b})Z$}& ATLAS 13 TeV \cite{69AZhatlas13}& 200-2000 & 3.2 fb$^{-1}$  \\
\hline
\end{tabular}
\end{footnotesize}
\caption{The upper limits at 95\%  C.L. on the production cross-section times branching ratio of the processes considered in the $ H $ and $ A $  searches at the LHC.}
\label{tabh}
\end{table}

\item[(5)] The non-observation of additional Higgs bosons. The
$\textsf{HiggsBounds}$ \cite{hb1,hb2} is used to implement the exclusion
constraints from the searches for the neutral and charged Higgs at LEP at 95\% confidence level.

At the LHC, the ATLAS and CMS have searched for additional scalar state via its decay into various SM channels and
some exotic decays. There are destructive interference contributions of $b$-quark
loop and top quark loop for $gg\to A$ production in type-II 2HDM. The cross section decreases with increasing
of $\tan\beta$, reaches the minimum value for the moderate value of
$\tan\beta$, and is dominated by the $b$-quark loop for enough large
value of $\tan\beta$. For $gg\to H$ production, the cross
section depends on $\sin(\beta-\alpha)$ in addition to $\tan\beta$
and $m_H$. We employ $\textsf{SusHi}$ \cite{sushi} to compute the cross sections for $H$ and $A$ in the
gluon fusion and $b\bar{b}$-associated production at NNLO in QCD. A complete list of the searches for
additional Higgs considered by us is summarized in 
Table \ref{tabh} where some channels are taken from Ref. \cite{1608.02573}. 
Our previous paper shows that $\tan\beta$ is nearly required to be larger than 3.0
in the case of the wrong sign Yukawa coupling of the 125 GeV Higgs \cite{1701.02678}.
For such large $\tan\beta$, $\sigma(gg\to A)$, Br$(A\to \gamma\gamma)$ and $\sigma(gg\to H)$ are sizably suppressed.
Therefore, the $A\to \gamma\gamma$ channels fail to constrain the parameter space.
In addition, considering that $m_H$ is fixed at a large value, $m_H=$ 600 GeV, the $H\to \gamma\gamma,~WW,~ZZ,~hh,~AZ$
channels can be safely ignored. 
In fact, in our previous paper \cite{1701.02678}, we included these channels, and found that these channels do not
impose constraints on the parameter space in the case of the wrong sign Yukawa coupling and $m_H=$ 600 GeV.

\item[(6)] The observables of DM. We employ $\textsf{
micrOMEGAs}$ \cite{micomega} to calculate the relic density and the today DM pair-annihilation cross sections.
The model file is generated by $\textsf{FeynRules}$ \cite{feyrule}.

For a small $m_S$, the $SS\to gg~,c\bar{c}~,\tau^+\tau^-$, $b\bar{b}$ annihilation channels play important
contributions to the DM relic density.
With increasing of $m_S$, the contributions of $SS\to WW,~ZZ,~hh,~t\bar{t}$ become important.
In addition to the annihilation into the SM particles, the DM can annihilate into $AA$, $HH$, $H^\pm H^\mp$, and $hH$  
if kinematically allowed.

In this model, the elastic scattering of $S$ on a nucleon receives
the contributions from the $h$ exchange diagrams. The spin-independent cross section is written as
\cite{sigis},
 \beq \sigma_{p(n)}=\frac{\mu_{p(n)}^{2}}{4\pi m_{S}^{2}}
    \left[f^{p(n)}\right]^{2},
\eeq where $\mu_{p(n)}=\frac{m_Sm_{p(n)}}{m_S+m_{p(n)}}$, \beq
f^{p(n)}=\sum_{q=u,d,s}f_{q}^{p(n)}\mathcal{C}_{S
q}\frac{m_{p(n)}}{m_{q}}+\frac{2}{27}f_{g}^{p(n)}\sum_{q=c,b,t}\mathcal{C}_{S
q}\frac{m_{p(n)}}{m_{q}},\label{fpn} \eeq with $\mathcal{C}_{S
q}=\frac{\lambda_h m_q}{m_h^2}y_q$. Following the recent study
\cite{1312.4951}, we take
\begin{eqnarray}
f_{u}^{p}\approx0.0208,\quad & f_{d}^{p}\approx0.0399,\quad &
f_{s}^{p}\approx0.0430,
\quad f_{g}^{p}\approx0.8963,\nonumber\\
f_{u}^{n}\approx0.0188,\quad & f_{d}^{n}\approx0.0440,\quad &
f_{s}^{n}\approx0.0430,\quad  f_{g}^{n}\approx0.8942.
 \label{eq:neuclon-form}
\end{eqnarray}
Where $f_{q}^{p}$ ($f_{q}^{n}$) is the form factor at the proton (neutron) for a light quark $q$,
and $f_{g}^{p}$ ($f_{g}^{n}$) is the form factor at the proton (neutron) for gluon.
If $f_{q}^{p}=f_{q}^{n}$ and
$f_{g}^{p}=f_{g}^{n}$ are satisfied, the $S$-nucleon scattering is
always isospin-conserving. If the relations are not satisfied, the $S$-nucleon scattering may be
isospin-violating for the appropriate values of $y_d$ and $y_u$.

Recently, the density of cold DM in the universe was estimated by the Planck
collaboration to be $\Omega_{c}h^2 = 0.1198 \pm 0.0015$ \cite{planck}. 
The strongest constraints on the spin-independent DM-nucleon cross section are 
from the PandaX-II (2017) for a DM with mass larger than 100 GeV \cite{pandax}, 
and from the XENON1T (2017) for a DM with mass smaller than 60 GeV \cite{xenon}.
For a DM with mass in the range of 60 GeV and 100 GeV, the upper limits of PandaX-II (2017) are
nearly the same as those of XENON1T (2017).
The Fermi-LAT search for the DM annihilation from dSphs gave the upper limits on the average cross sections
of the DM annihilation into $e^+ e^-$, $\mu^+ \mu^-$, $\tau^+\tau^-$, $u\bar{u}$, $b\bar{b}$, and $WW$ \cite{fermi}.

\end{itemize}

\section{Results and discussions}
\begin{figure}[tb]
 \epsfig{file=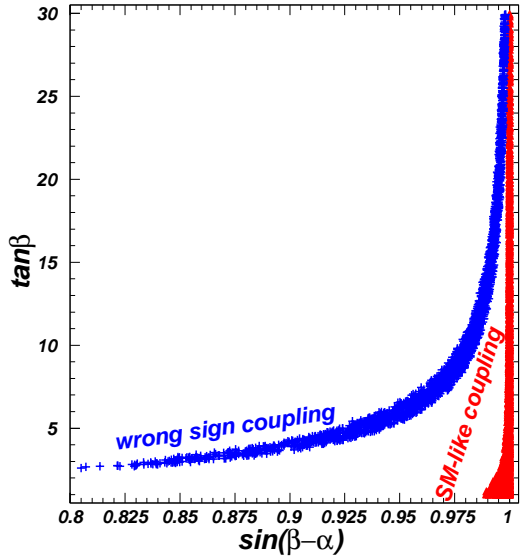,height=7.5cm}
\vspace{-0.5cm} \caption{The samples surviving from the constraints of the 125 GeV Higgs signal data
projected on the plane of $\sin(\beta-\alpha)$ versus $\tan\beta$.} \label{athe1}
\end{figure}

\begin{figure}[tb]
  \epsfig{file=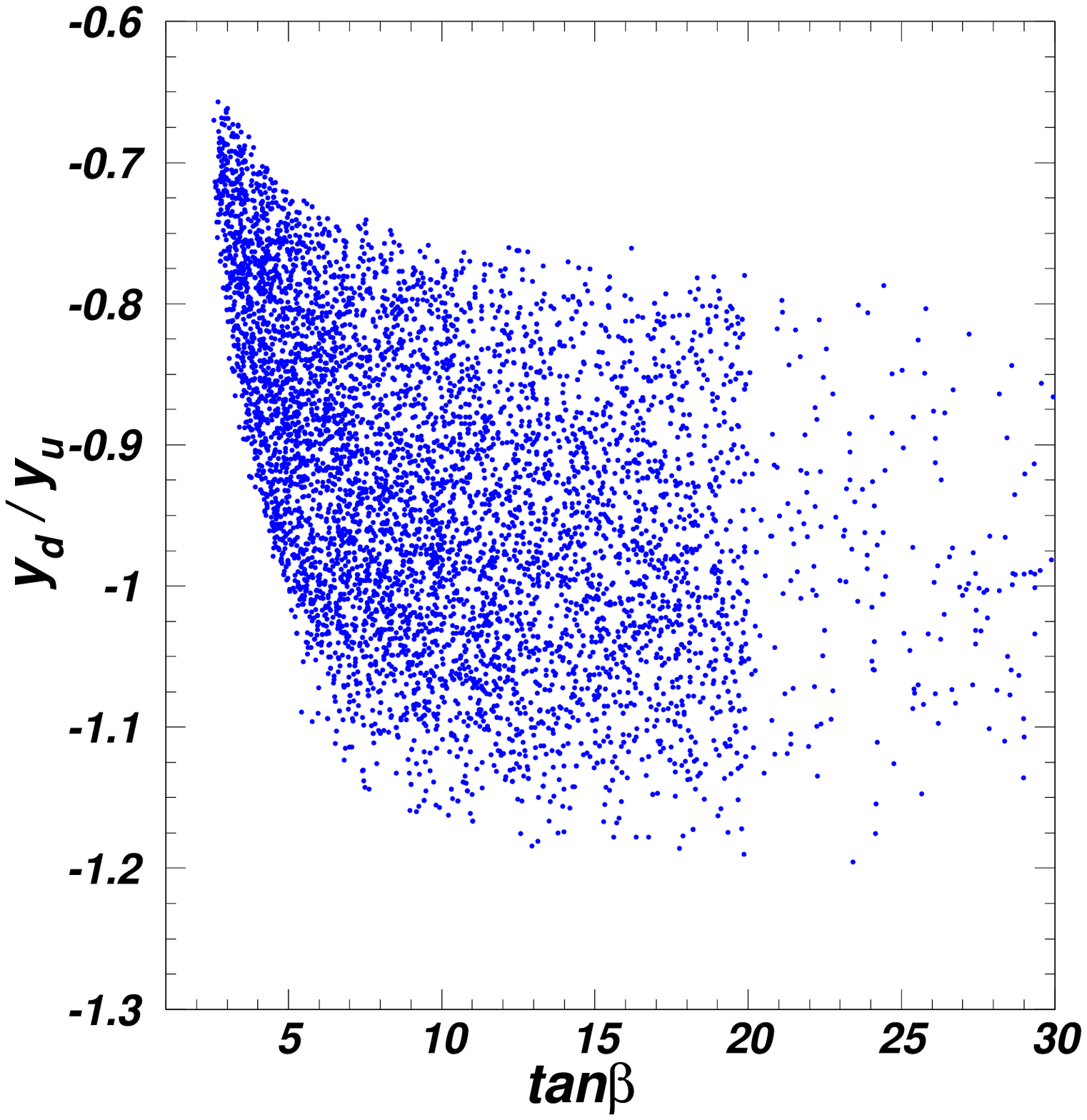,height=7.5cm}
 \epsfig{file=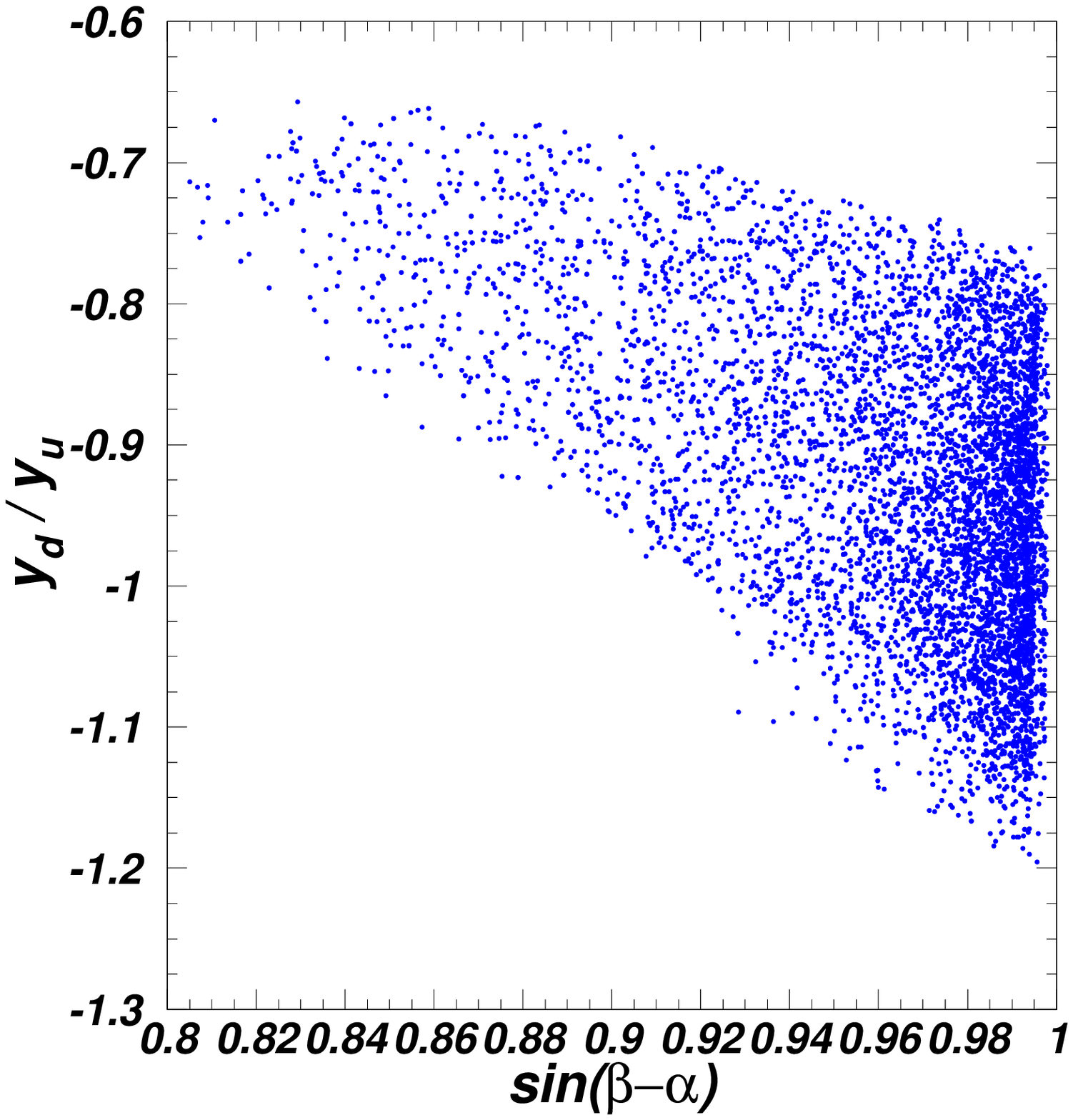,height=7.5cm}
\vspace{-0.5cm} \caption{In the case of the wrong sign Yukawa coupling of the 125 GeV Higgs, 
the samples surviving from the constraints of the 125 GeV Higgs signal data projected on the
planes of $y_d/y_u$ versus $\tan\beta$ and $y_d/y_u$ versus $\sin(\beta-\alpha)$. $y_d$
 ($y_u$) denotes the normalized factor of down-type (up-type) quark Yukawa coupling of the 125 GeV Higgs
with respect to the SM.}\label{ydyu}
\end{figure}

In Fig. \ref{athe1}, we show $\sin(\beta-\alpha)$ and $\tan\beta$ allowed by the signal data of the
125 GeV Higgs. Fig. \ref{athe1} shows that $\sin(\beta-\alpha)$ is required to be very close to 1 in the case
of the SM-like Higgs coupling of the 125 GeV Higgs.
However, in the case of the wrong sign Yukawa coupling of the 125 GeV Higgs, $\sin(\beta-\alpha)$ is allowed to be much smaller than
1, and $\tan\beta$ is restricted to a very narrow range for
a given value of $\sin(\beta-\alpha)$, such as $5.4<\tan\beta<6.6$ for
$\sin(\beta-\alpha)=0.95$. The surviving samples with the wrong sign Yukawa coupling are projected on the planes
of $y_d/y_u$ versus $\tan\beta$ and $y_d/y_u$ versus $\sin(\beta-\alpha)$ in Fig. \ref{ydyu}.
As discussed above, the down-type quark Yukawa coupling of the 125 GeV Higgs has an opposite sign to 
the Yukawa coupling of up-type quark. $y_d/y_u$ is allowed to vary from -1.2 to -0.65, and 
the absolute value increases with $\tan\beta$ and $\sin(\beta-\alpha)$.

\begin{figure}[tb]
  \epsfig{file=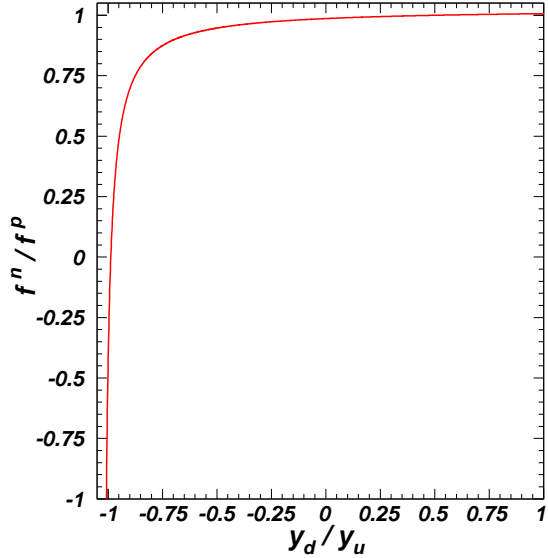,height=7.5cm}
\vspace{-0.5cm} \caption{$f^n/f^p$ versus $y_d/y_u$.} \label{isospin}
\end{figure}

\begin{figure}[tb]
   \epsfig{file=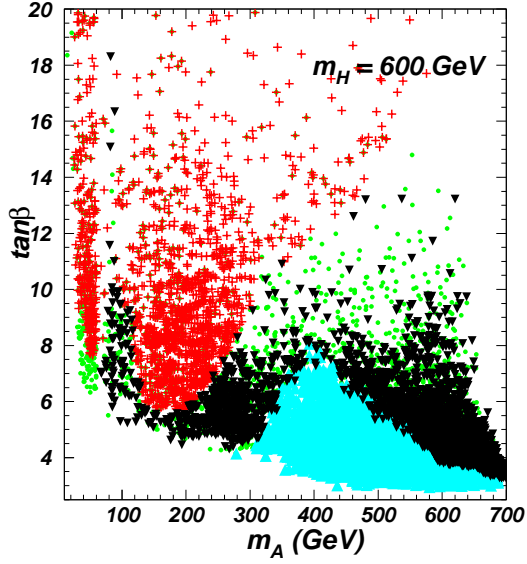,height=7.5cm}
\vspace{-0.5cm}
   \caption{In the wrong sign Yukawa coupling of the 125 GeV Higgs, the surviving samples projected on
the plane of $\tan\beta$ versus $m_A$. All the samples are allowed by the constraints of "pre-LHC" and the
125 GeV Higgs signal data. Also the constraints of the DM relic density and the searches for Higgs at LHC
are satisfied for the inverted triangles (black). The pluses (red) and triangles (sky blue) are
respectively excluded by the $A/H\to \tau^+ \tau^-$ and $A\to hZ$ searches at LHC. } \label{lhc}
 \end{figure}

\begin{figure}[tb]
  \epsfig{file=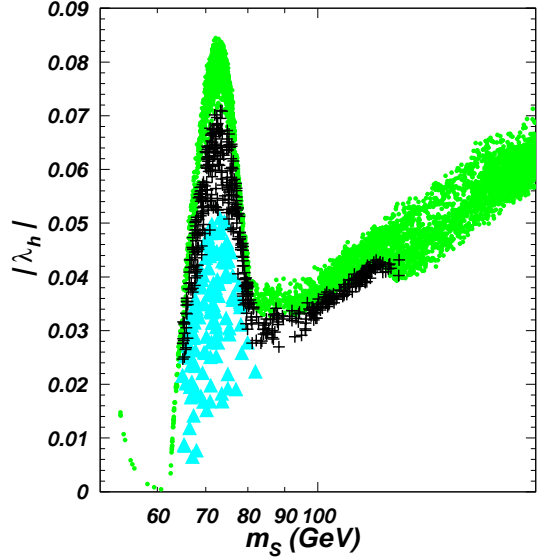,height=7.5cm}
\vspace{-0.5cm}
  \caption{The surviving samples projected on the plane of $\mid \lambda_h\mid$ versus $m_S$.
All the samples are allowed by the constraints of "pre-$\Omega_c h^2$" and the DM relic density.
The contribution of $SS\to AA$ to $1/(\Omega h^2)$ is $0\sim 10\%$ for the bullets (green), 
$10\%\sim50\%$ for the pluses (black), and $50\%\sim 100\%$ for triangles (sky blue).} \label{relic}
\end{figure}

\begin{figure}[tb]
  \epsfig{file=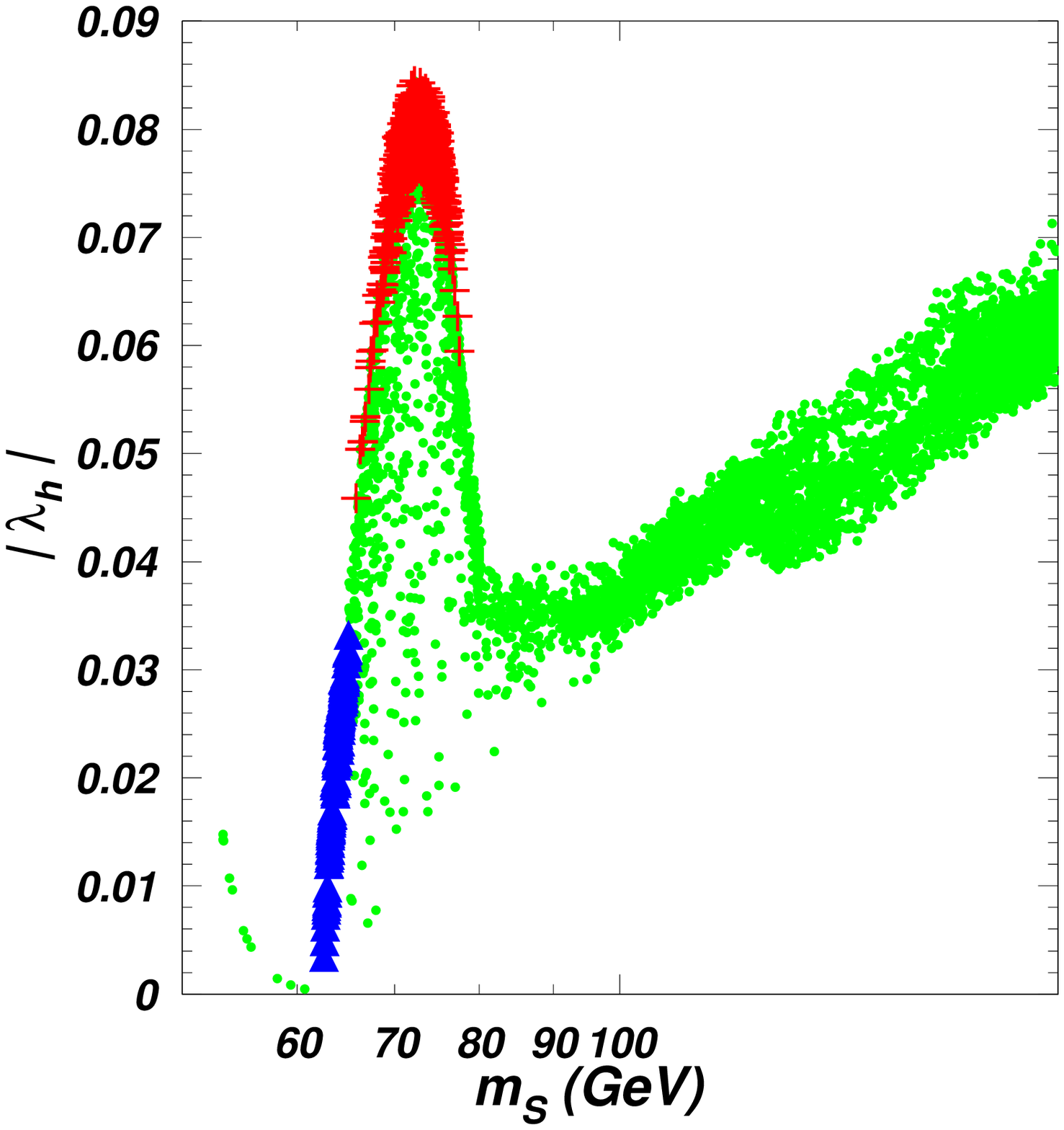,height=7.58cm}
\epsfig{file=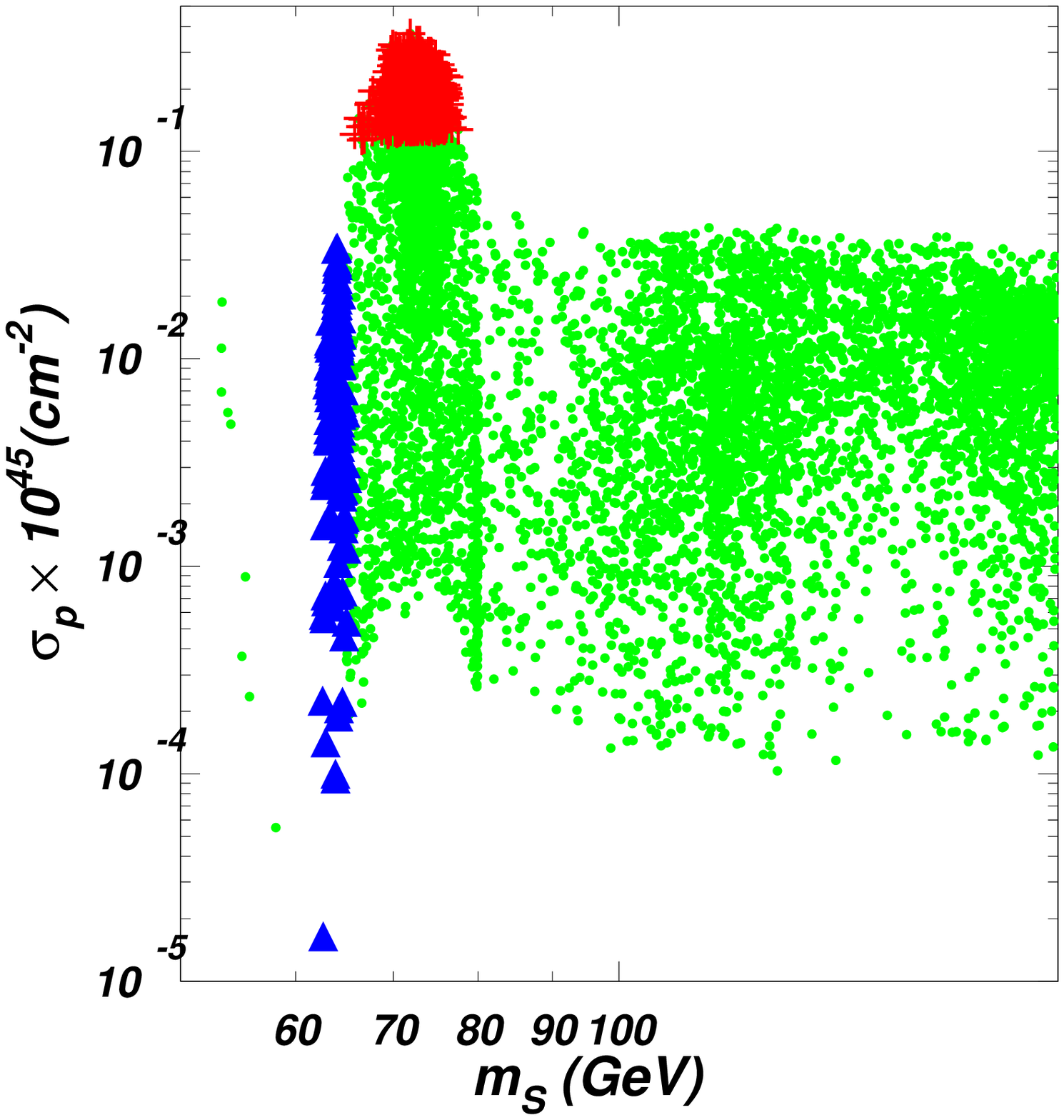,height=7.5cm}
\vspace{-0.5cm}
  \caption{The surviving samples projected on the planes of $\mid \lambda_h\mid$ versus $m_S$
and $\sigma_p$ versus $m_S$. All the samples are allowed by the constraints of "pre-$\Omega_c h^2$" and the DM relic density.
The pluses (red) are excluded by the constraints on the spin-independent DM-proton cross section 
from XENON1T (2017) and PandaX-II (2017). 
The triangles (royal blue) are excluded by the Fermi-LAT search for DM annihilation from dSphs.} \label{sigmap}
\end{figure}

\begin{figure}[tb]
 \epsfig{file=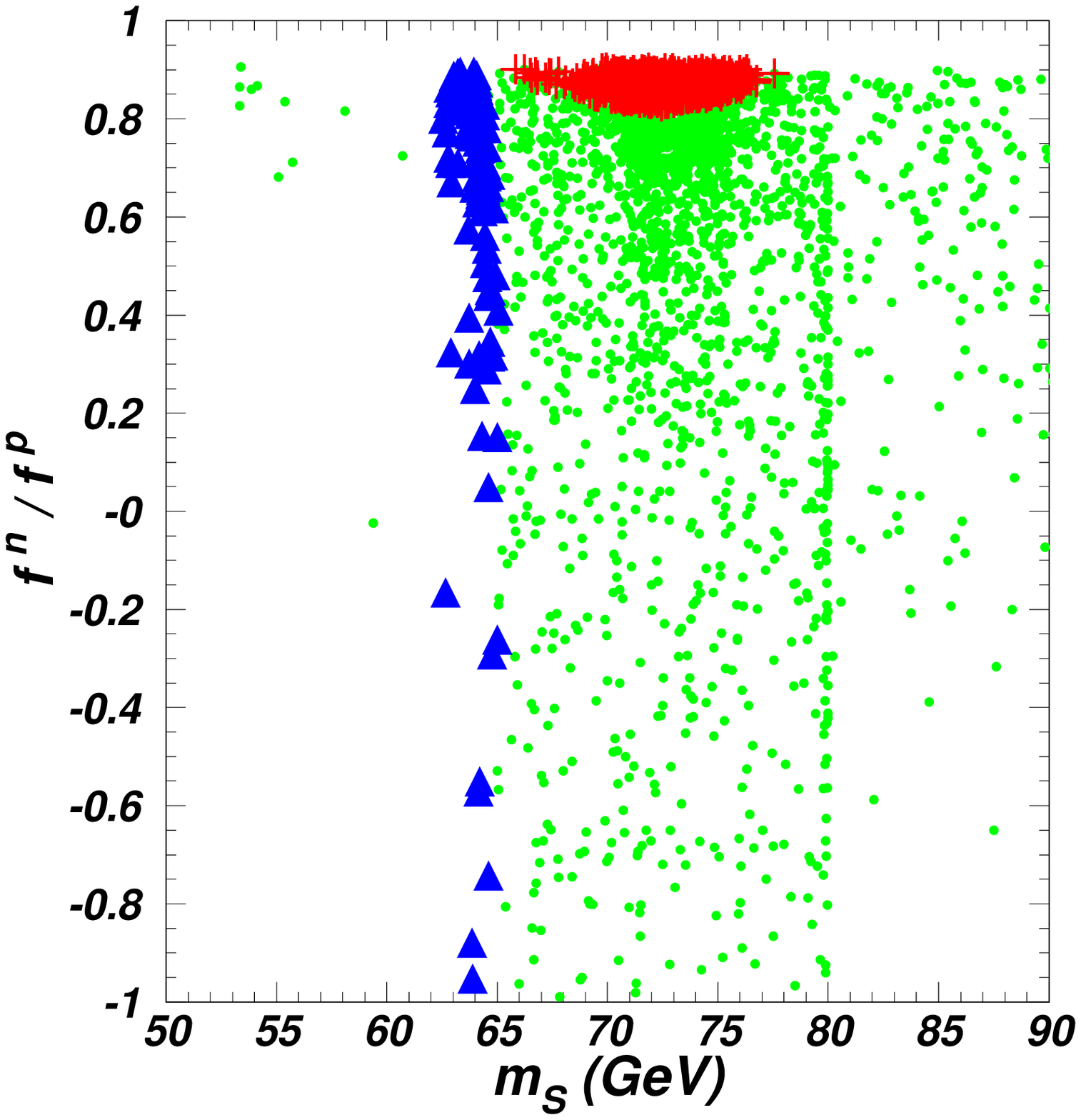,height=5.60cm}
  \epsfig{file=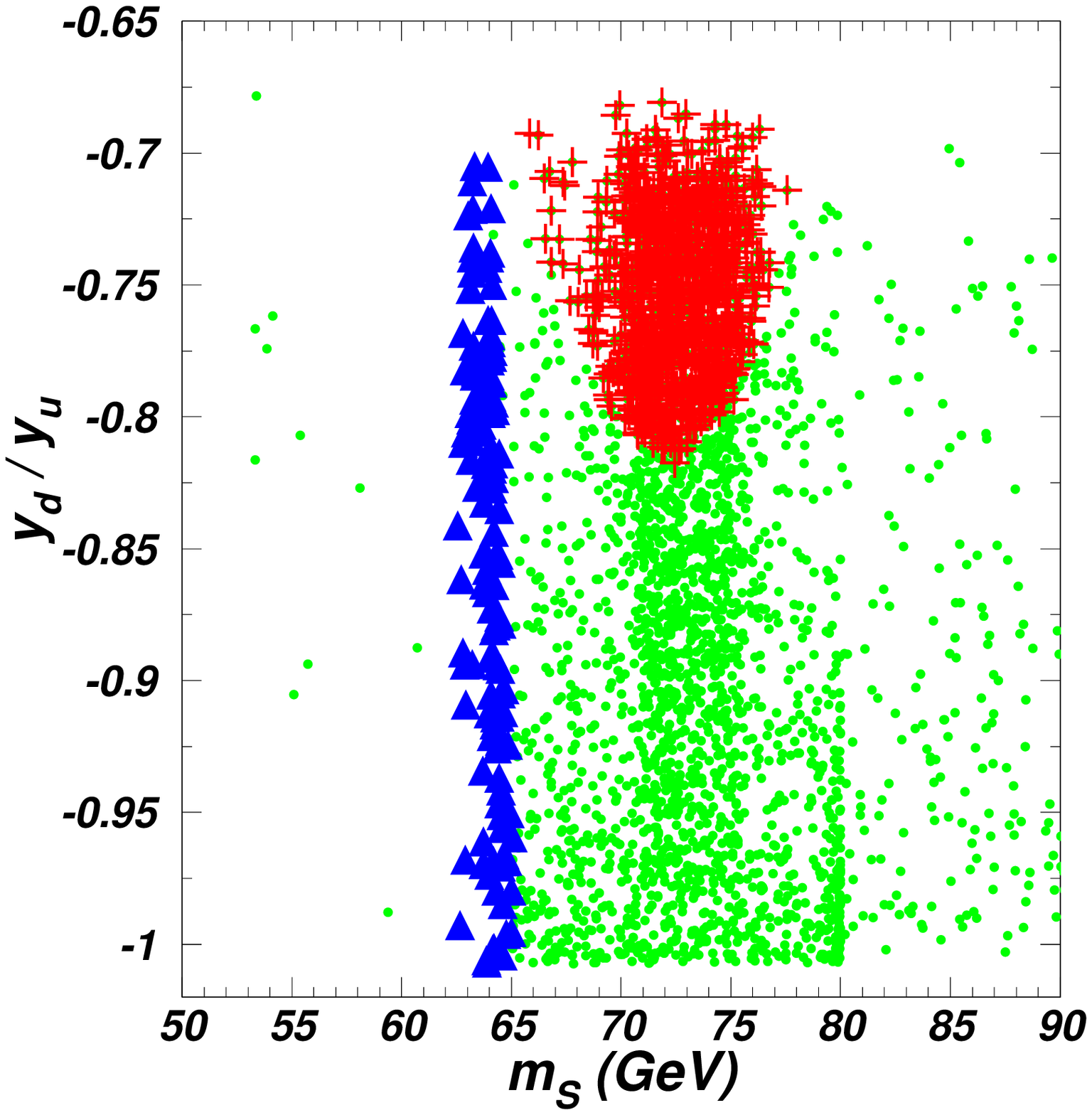,height=5.60cm}
 \epsfig{file=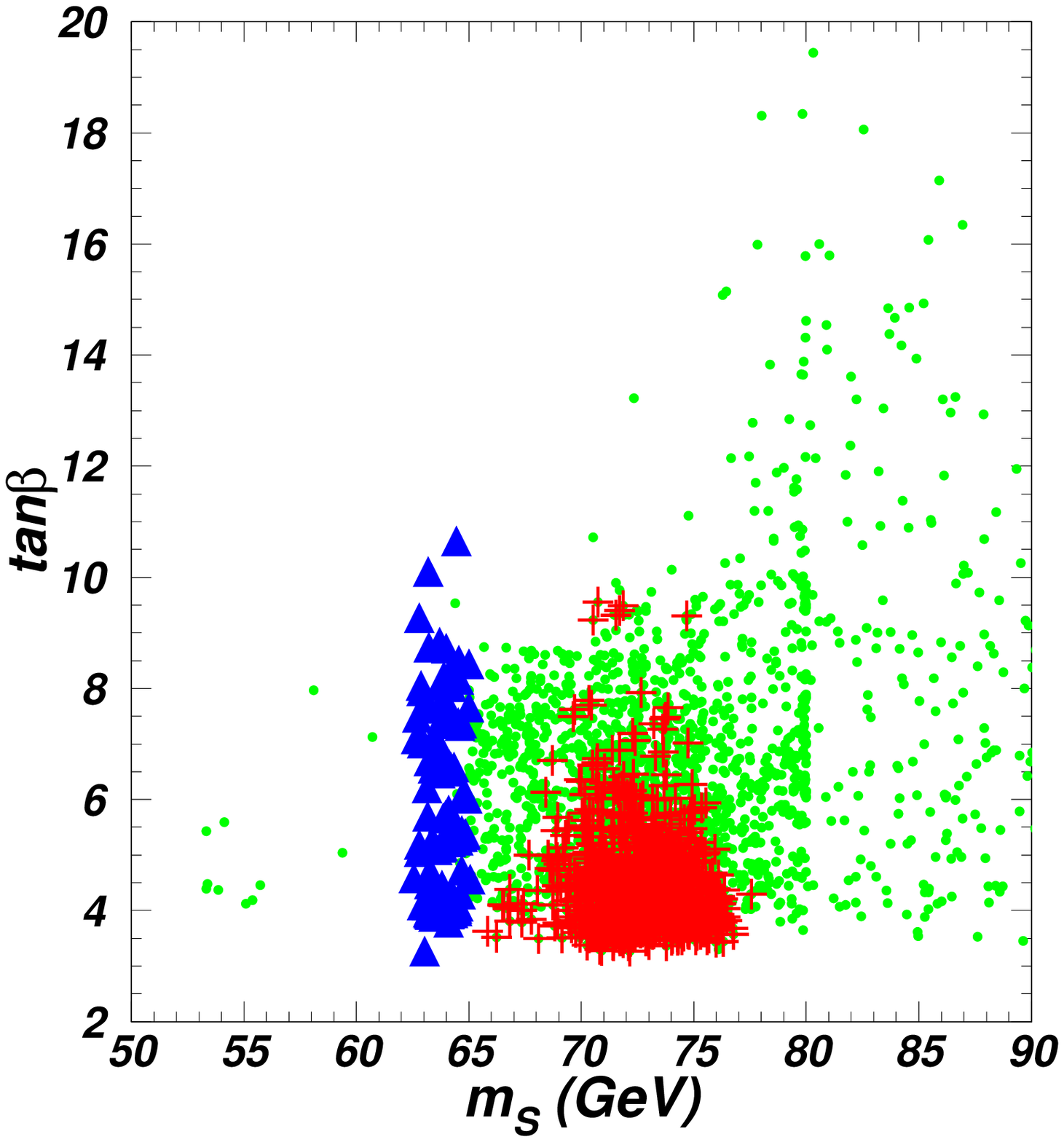,height=5.60cm}
\vspace{-0.5cm} \caption{Same as Fig. \ref{sigmap}, but projected on the planes of 
$f^n/f^p$ versus $m_S$, $y_d/y_u$ versus $m_S$, and $\tan\beta$ versus $m_S$.} \label{sigmap2}
\end{figure}

In Fig. \ref{isospin}, we show $f^n/f^p$ versus $y_d/y_u$.
Since the hadronic quantities in the spin-independent DM-nucleon
scattering are fixed, $f^n/f^p$ only depends on the normalized
factors of Yukawa couplings, $y_u$ and $y_d$. The $f^n/f^p$ is very sensitive
to $y_d/y_u$ for $y_d/y_u$ around -1.0, and very close to 1.0 for
$y_d/y_u>0$. In the following discussions, we will focus on the
surviving samples with $-1.0< f^n/f^p<1.0$ where the upper limits on the spin-independent DM-nucleon cross section
from the XENON1T (2017) and PandaX-II (2017) can be weakened.

In Fig. \ref{lhc}, we project the surviving samples on the
planes of $\tan\beta$ versus $m_A$ after imposing the constraints of "pre-LHC"
(denoting the theory, the oblique parameters, the flavor observables, $R_b$,
 and the exclusion limits from searches for Higgs at LEP), the signal data of the 125 GeV Higgs,
the searches for the additional Higgs at LHC, and the DM relic density.
Since the signal data of the 125 GeV Higgs impose very strong constraints on the width of $(h\to AA)$,
the $h\to AA$ channels at the LHC fail to give constraints on the parameter space.
The $AhZ$ coupling is proportional to $\cos(\beta-\alpha)$, and $\sin(\beta-\alpha)$ increases with $\tan\beta$.
Therefore, the $A\to hZ$ channel can impose a lower bound on $\tan\beta$ for $m_A>$ 280 GeV. 
While the $b\bar{b}\to A \to \tau^+ \tau^-$ channel can impose an upper bound on $\tan\beta$. 
For 140 GeV $<m_A<200$ GeV, $\tan\beta$ is restricted to a very narrow range, $4.2<\tan\beta<5.7$.
However, for $m_A$ around 80 GeV, $\tan\beta$ is allowed to be as large as 18. 
The correct DM relic density can be obtained for the nearly whole parameter space of $m_A$ and $\tan\beta$.
Certainly, the relic density is sensitive to the DM mass and the DM coupling with the 125 GeV Higgs.

In Fig. \ref{relic}, we project the surviving samples on the
plane of $\mid\lambda_h\mid$ versus $m_S$ after imposing the constraints of the DM relic density and  
"pre-$\Omega_{c}h^2$" (denoting "pre-LHC", the 125 GeV Higgs signal data, and the searches for Higgs at LHC).
From Fig. \ref{relic}, we find the $SS\to AA$ annihilation channel can play an important contribution to 
the DM relic density, especially for 65 GeV $<m_S<85$ GeV. In such range, the DM mass deviates from the resonance region,
and the $SS\to WW^{(*)}$ channel is kinematically suppressed. Therefore, the contribution of $SS\to AA$ to 
the relic density can be dominant.
Due to the contribution of $SS\to AA$ channel to the relic density, the DM coupling with
the 125 GeV Higgs can be sizably suppressed. For example, when 
the contribution of $SS\to AA$ to $1/(\Omega h^2)$ is smaller than $10\%$, $\mid \lambda_h\mid$ is required to
be in the range of 0.065 and 0.08 for $m_S=$ 70 GeV. When the contribution of $SS\to AA$ is dominant, $\mid \lambda_h\mid$ is
allowed to be as low as 0.015 for $m_S=$ 70 GeV. However, for $m_S < 60$ GeV, the $SS\to AA$ annihilation channel can not
give sizable contribution to the relic density. For such case, the $h\to AA$ decay mode will open, and
enhance the total width of the 125 GeV Higgs. The signal data of the 125 GeV Higgs will give   
strong constraints on the $hAA$ coupling in addition to the $hSS$ coupling. Due to the tension between 
the DM relic density and the signal data of the 125 GeV Higgs, $m_S<50$ GeV is excluded.
Besides, a very small $\mid\lambda_h\mid$ still can achieve the correct
relic abundance at the resonance, $m_S$ around 60 GeV.

In Fig. \ref{sigmap} and Fig. \ref{sigmap2}, we project the surviving samples on the planes of 
$\mid\lambda_h\mid$, $\sigma_p$, $f^n/f^p$, $y_d/y_u$, and $\tan\beta$ versus $m_S$
after imposing the constraints of "pre-$\Omega_{c}h^2$", the DM relic density, XENON1T (2017), PandaX-II (2017), 
and the Fermi-LAT search for DM annihilation from dSphs.
The Fig. \ref{sigmap} shows that for 65 GeV $<m_S<$ 78 GeV, 
the relic density allows $\mid\lambda_h\mid$ to have a relatively
large value, which can lead to a large spin-independent $\sigma_p$. 
In the range of 65 GeV $<m_S<$ 78 GeV, the upper limits of XENON1T (2017) and PandaX-II (2017) exclude 
$\sigma_p>1\times 10^{-46}~cm^{-2}$. In the range of 65 GeV $<m_S<$ 78 GeV, there are many samples surviving from
the constraints of XENON1T (2017) and PandaX-II (2017), especially for the case that $SS\to AA$ plays an important contribution
to the relic density. For such case, $\lambda_h$ can be sizably suppressed,  leading $\sigma_p$ to accommodate the upper
limits of XENON1T (2017) and PandaX-II (2017). 
The Fermi-LAT limits can exclude most of samples in the range of 62.5 GeV $<m_S <65$ GeV, even including
the sample with $\sigma_p \sim 10^{-50}~cm^{-2}$  and $m_S\sim 62.5$ GeV. For such sample, the DM pair-annihilation 
can be sizably enhanced at the resonance, but the spin-independent DM-nucleon cross section has no resonance enhancement.
When $m_S$ is moderately smaller than $m_h/2$,
a very small $\mid\lambda_h\mid$ can achieve the correct
relic abundance since the integral in the calculation of thermal
average can be dominated by the resonance at $s = m_h^2$ with $s$ being the squared
center-of-mass energy of the pair-annihilation of DM.
For such case, the today average cross section of DM pair-annihilation can be sizably suppressed 
since the velocity of DM at the present time is much smaller than that in the early universe.
Therefore, the limits of Fermi-LAT search for DM annihilation from dSphs can be
satisfied when $m_S$ is moderately smaller than $m_h/2$.

Fig. \ref{sigmap2} shows that the limits of XENON1T (2017) and PandaX-II (2017) can be satisfied
for $-1<f^n/f^p <$ 0.8 and $y_d/y_u <$ -0.82. The DM scattering rate with Xe target can be sizably
suppressed for $f^n/f^p$ around -0.7, thus weakening the constraints from XENON1T (2017) and PandaX-II. 
However, for $f^n/f^p$ around -0.7, the Fermi-LAT still can exclude the samples in the range of 62.5 GeV $<m_S<$ 65 GeV.
The right panel shows that most of samples excluded by the XENON1T (2017) and PandaX-II (2017) 
lie in the ranges of $3<\tan\beta <8$ and 65 GeV $<m_S<$ 78 GeV.

\section{Conclusion}
The wrong sign Yukawa coupling of the 125 GeV Higgs is an interesting characteristic of 2HDM.
Such a scalar is taken as the portal between the DM and SM sectors, the isospin-violating DM interactions with nucleons
can be realized. We examine the 125 GeV Higgs with the wrong sign Yukawa coupling of the down-type
quark which is the only portal between the 
DM and SM sectors in the framework of the type-II 2HDM with a scalar DM.
 After imposing the constraints from the theory, oblique parameters, the flavor observables, the Higgs searches at the LHC,
and the DM experiments, we obtain some interesting observables:
(i) In the case of the wrong sign Yukawa coupling of the 125 GeV Higgs, the searches for additional Higgs
via $\tau^+\tau^-$ channel at the LHC
can impose an upper limit on $\tan\beta$, and the $A\to hZ$ channel can impose a lower limit on $\tan\beta$.
(ii) The $SS\to AA$ annihilation channel can play an important contribution to 
the DM relic density, especially for 65 GeV $<m_S<85$ GeV. Due to the tension between 
the DM relic density and the signal data of the 125 GeV Higgs, $m_S<50$ GeV is excluded.
(iii) The upper limits of XENON1T (2017) and PandaX-II (2017) on the spin-independent DM-nucleon cross section exclude 
most of samples in the ranges of 65 GeV $<m_S<$ 78 GeV, $0.8<f^n/f^p<1$, and $y_d/y_u>-0.82$.
(iv) The Fermi-LAT limits can exclude most of samples in the range of 62.5 GeV $<m_S <65$ GeV.
For $m_S$ around $m_h/2$, the sample with $f^n/f^p\sim -0.7$ can still be excluded
by the Fermi-LAT limits. We simply summarize the ranges of $m_S$ and several corresponding key parameters 
surviving from all bounds in Table \ref{dmmass}. For $m_S>$ 200 GeV, the limits of XENON1T (2017), PandaX-II (2017),
and the Fermi-LAT fail to constrain the parameter space, the model can simply satisfy the limits of 
the relic density and Higgs searches at the LHC. The relevant discussions are shown in detail in Ref. \cite{2hisos-6}, and we
do not include the scenario of $m_S>$ 200 GeV in this paper.
\begin{table}
\begin{footnotesize}
\begin{tabular}{| c | c | c | c |}
\hline
\textbf{$m_S$} & \textbf{$\mid \lambda_h\mid$}~(depending on $m_S$) & \textbf{$y_d/y_u$}  &  \textbf{$f^n/f^p$} \\
\hline
 50 GeV $\sim$ 62.5 GeV & $4\times 10^{-4}\sim$ 0.016               & $-1.01\sim -0.65$ & $-1.0\sim 1.0$ \\
 65 GeV $\sim$ 78 GeV   &  $0.008\sim 0.076$                        & $-1.01\sim -0.82$ & $-1.0\sim 0.8$ \\
 78 GeV $\sim$ 200 GeV  &  $0.02\sim 0.072$                         & $-1.01\sim -0.7$ & $-1.0\sim 0.9$ \\
\hline
\end{tabular}
\end{footnotesize}
\caption{The ranges of $m_S$ and several corresponding key parameters 
surviving from all bounds.}
\label{dmmass}
\end{table}

If the 125 GeV Higgs with the SM-like coupling mediates the DM interactions
with SM particles, the LUX and PandaX-II limits exclude the DM mass up to 330 GeV, except a small range near
the resonance point $m_{DM}=m_h/2$ in the case of DM only annihilation into the SM particles \cite{2hisos-6}.
However, in the case of the wrong sign Yukawa coupling of the 125 GeV Higgs, the more stringent limits
from PandaX-II (2017) and XENON1T (2017) still allow the DM mass to be as low as 50 GeV for the appropriate isospin-violating DM interactions 
with nucleons.

\section*{Acknowledgment}
We thank Prof. Junjie Cao for helpful discussions.
 This work is supported by the National Natural Science Foundation
of China under grant No. 11575152, and the Natural Science Foundation of
Shandong province (ZR2017JL002 and ZR2017MA004).

\end{document}